\documentclass[aps,prl,letter,twocolumn,showpacs,superscriptaddress]{revtex4-1}

\usepackage{amsmath}
\usepackage{amsfonts}
\usepackage{amssymb}
\usepackage{dcolumn}
\usepackage{graphicx}
\usepackage{color}
\usepackage{bm}

\def\dd{{\mathrm d}}
\def\ii{{\mathrm i}}
\def\ee{{\mathrm e}}
\def\K{{\mathrm K}}

\begin{document}

\title{Enhancement of Blackbody Friction due to the Finite Lifetime
of Atomic Levels}

\author{G. \L{}ach\email{gel@physi.uni-heidelberg.de}}
\author{M. DeKieviet\email{maarten@physi.uni-heidelberg.de}}

\affiliation{Physikalisches Institut der Universit\"{a}t, 
Albert-Ueberle-Strasse 3-5, 69120 Heidelberg, Germany}

\author{U. D. Jentschura\email{ulj@mst.edu}}

\affiliation{Department of Physics, Missouri University of Science and Technology, 
Rolla, Missouri 65409-0640, USA}

\begin{abstract}
The thermal friction force acting on an atom
moving relative to a thermal photon bath 
is known to be proportional to an integral over the 
imaginary part of the frequency-dependent atomic (dipole) polarizability. 
Using a numerical approach, we find that blackbody friction on
atoms either in dilute environments or in hot ovens is larger 
than previously thought by orders of magnitude.  This enhancement is due to far
off-resonant driving of transitions by low-frequency thermal radiation.
At typical temperatures, the blackbody radiation maximum lies far below the
atomic transition wavelengths. Surprisingly, 
due to the finite lifetime of atomic levels, which gives rise to Lorentzian line
profiles, far off-resonant excitation leads to the dominant
contribution to the blackbody friction.
\end{abstract}

\pacs{68.35.Af, 12.20.Ds, 95.30.Dr, 95.30.Jx}

\maketitle

{\em Introduction.---}In Ref.~\cite{MkPaPoSa2003}, the thermal drag force on an
atom moving through a thermal bath at velocity $v$ has been calculated on the
basis of the fluctuation-dissipation theorem.  In a nutshell, the
fluctuation-dissipation theorem states that any thermal fluctuation of a
physical quantity (say, the electric field at finite temperature) is
accompanied by corresponding fluctuations in the conjugate variable (here, the
atomic dipole moment) provided the susceptibility (in the current case, the
atomic polarizability) has a nonvanishing imaginary part. The imaginary part
describes a dissipative process, in which the atom absorbs, then spontaneously
emits, electromagnetic radiation. The dissipative fluctuations give rise to a
drag force calculated using the Green--Kubo formula, as thoroughly explained in
Ref.~\cite{MkPaPoSa2003}.  Further physical insight can be gained if one
understands the process in terms of the direction-dependent Doppler
effect~\cite{MK1979bb}. The atom absorbs blue-shifted blackbody photons
coming in from the front, while emitting these photons in all directions,
thereby losing kinetic energy due to net drain on its energy, and, as
a consequence, on its momentum~\cite{LaDKJe2011pra}.

Here, we show that even more intriguing problems arise when one tries to
evaluate the effect numerically, for simple atoms. In atomic physics, the width
$\Gamma_n$ for each individual energy level $n$ needs to be 
determined separately. Atomic transitions can be driven
even very far from resonance, albeit with small transition probabilities,
The blackbody spectrum is distributed over the entire frequency interval
$\omega \in [0, \infty)$, which leads to significant non-resonant contributions to the
thermal friction. 

The authors of Ref.~\cite{MkPaPoSa2003} use correlation functions for the
thermal electromagnetic fluctuations~\cite{RyKrTa1989vol3,PiLi1958vol9}, in order to
calculate the friction force acting on neutral, polarizable objects moving
through uniform and isotropic thermal radiation.  
According to Eq.~(12) of Ref.~\cite{MkPaPoSa2003}, the effective
friction (EF) force, which acts in a direction opposite to the velocity $v$, is
given as a spectral integral,
\begin{equation}
\label{force1}
F_{\rm EF} = -\frac{\beta\hbar^2\,v}{3\pi\,c^5\,(4 \pi \epsilon_0)}
\int_0^{\infty}\dd\omega\,
\frac{\omega^5\, {\rm Im}\,\alpha(\omega)}
{\sinh^2(\tfrac12 \beta \hbar \omega)} \,,
\end{equation}
where $\beta = 1/(k_B T)$ is the Boltzmann factor and $\alpha(\omega)$
is the dynamic polarizability of the atom. 
We here argue that the inclusion
of the resonance widths due to the finite lifetimes of atomic levels
is crucial in the calculation of the friction force.
SI mksA units are used throughout this work.

{\em Narrow and finite width.---} If we assume
that all atomic transitions are infinitely narrow
(of width $\epsilon$), then
\begin{equation}
\alpha(\omega) =
\sum_n \frac{f_{0n}}{2\omega_{0n}}
\left(
\frac{1}{\omega_{0n} - \omega - \ii \epsilon} +
\frac{1}{\omega_{0n} + \omega - \ii \epsilon} 
\right)\,,
\end{equation}
where $f_{0n}$ denotes the oscillator strength of the transition
and $\omega_{0n}$ is the angular frequency for the 
transition from the ground state $|0\rangle$ to the
excited state $|n\rangle$. In view of the Dirac prescription
$1/(x - \ii \epsilon) = P(1/x) + \ii \pi \delta(x)$, 
the imaginary part of the polarizability
is approximated as a sum of Dirac $\delta$ peaks,
\begin{equation}
\label{delta}
{\rm Im}\,\alpha(\omega) =
\sum_n \frac{\pi f_{0n}}{2\omega_{0n}}\,\delta(\omega-\omega_{0n})\,.
\end{equation}
However, if one 
includes the width $\Gamma_n$ of the excited states, then 
the starting expression (see Chap.~8 of Ref.~\cite{Lo1993})
for the dynamic polarizability reads
%
\begin{align}
\label{20}
\alpha(\omega) =& \;
\sum_n 
\sum_{\pm}
\frac{f_{0n}}{2\omega_{0n}} \;
\frac{1}{\omega_{0n} - \frac{\ii}{2} \Gamma_n(\omega) \pm \omega} \,,
\nonumber\\
{\rm Im}\,\alpha(\omega) =& \;
\sum_n \frac{f_{0n}\, \Gamma_n(\omega)\;\omega_{0n}}%
{\big(\omega^2-\omega_{0n}^2\big)^2+
\omega_{0n}^2\, \Gamma_n^2(\omega)} \,.
\end{align}
%
Here, the decay width $\Gamma_n(\omega)$ may be
a function of the driving frequency $\omega$.
In a number of places in the literature [e.g., see the text 
after Eq.~(2) of Ref.~\cite{LaSoPlSo2002}], it is assumed that
\begin{align}
\label{ansatz1}
\Gamma_n(\omega) = \overline \Gamma_n(\omega) = 
\frac{\omega}{\omega_{0n}} \Gamma_n \,,
\qquad
\Gamma_n \equiv \Gamma_n(\omega_{0n}) \,.
\end{align}
One can justify  the ansatz~\eqref{ansatz1} in two ways (i)~and(ii).
(i)~One may invoke an analogy
with a damped, driven harmonic oscillator, whose Green function
$g(t - t')$ fulfills the defining differential equation
\begin{align}
& \left( - \frac{\partial^2}{\partial t^2}
+ \gamma \, \frac{\partial}{\partial t}
+ \omega_0^2 \right) \overline g(t - t') = \delta(t-t) \,,
\end{align}
so that the Fourier transform of the Green function reads as
$\overline g(\omega) = 1/(\omega_0^2 - \omega^2 - \ii \, \gamma \, \omega)$,
with
\begin{equation}
\label{img}
{\rm Im} \; \overline g(\omega) =
\frac{\gamma \, \omega}{ \left( \omega^2 -  \omega_0^2 \right)^2 +
\gamma^2 \, \omega^2} \,.
\end{equation}
Assuming that $\Gamma_n(\omega) = \overline \Gamma_n(\omega)$,
this is proportional to the expression 
in~\eqref{20} under the obvious identification
$\omega_0 \to \omega_{0n}$, $\gamma \to \Gamma_n$.
(ii)~The decay width $\Gamma_n$ enters 
the propagator denominators in Eq.~\eqref{20}
by a summation of self-energy insertions~\cite{BeSa1957}.
The imaginary of the self-energy, divided by $\hbar$,
equals the decay width~\cite{BaSu1978}. 
The velocity-gauge expression~\cite{BeSa1957,Sa1994Mod,Sa1967Adv} 
for the decay rate, at resonance $\omega = \omega_n$ and off resonance
(for general $\omega$), reads as
\begin{subequations}
\label{velgauge}
\begin{align}
\label{velgauge1}
\Gamma_n =& \; 
\frac{4 \alpha}{3 \pi} \, \omega_{0n} \, 
\frac{\left| \left< \Psi_0 | \, \vec p \, | \Psi_n \right> \right|^2}{(mc)^2} \,,
\\[0.23ex]
\label{velgauge2}
\overline \Gamma_n(\omega) =& \; \frac{4 \alpha}{3 \pi} \, \omega
\, \frac{\left| \left< \Psi_0 | \, \vec p \, | \Psi_n \right> \right|^2}{(mc)^2} =
\frac{\omega}{\omega_{0n}} \Gamma_n\,,
\end{align}
\end{subequations}
where $\vec p$ is the momentum operator,
and $\Psi_0$ and $\Psi_n$ are the wave functions of the
ground and excited state.

By contrast, 
the so-called length gauge expression~\cite{BeSa1957,Sa1994Mod,Sa1967Adv}
for the decay width off resonance reads as
\begin{subequations}
\label{lengauge}
\begin{align}
\label{lengauge1}
\Gamma_n =& \; \frac{4 \alpha}{3 \pi} \, \omega_{0n}^3 \, 
\frac{\left| \left< \Psi_0 | \vec x | \Psi_n \right> \right|^2}{c^2}\,,
\\[0.23ex]
\label{lengauge2}
\widetilde \Gamma_n(\omega) =& \; \frac{4 \alpha}{3 \pi} \, \omega^3
\, \frac{\left| \left< \Psi_0 | \vec x | \Psi_n \right> \right|^2}{c^2} =
\left( \frac{\omega}{\omega_{0n}} \right)^3 \, \Gamma_n\,.
\end{align}
\end{subequations}
For atoms, using the commutator relation 
$\vec p = \ii m [H, \vec x]/\hbar$, where $H$ is the Hamiltonian
and $m$ the electron mass,
one can show the equivalence of Eqs.~\eqref{velgauge1} and~\eqref{lengauge1}
at resonance. The $\omega^3$ dependence 
off resonance in length gauge can be justified 
by analogy with Abraham--Lorentz radiative damping,
with a damped oscillator Green function
\begin{equation}
\label{ablo}
\begin{split}
\left( - \frac{\partial^2}{\partial t^2}
- \frac{\gamma}{\omega_0^2} \, \frac{\partial^3}{\partial t^3}
+ \omega_0^2 \right) \widetilde g(t - t') = \delta(t-t)\,,
\\
{\rm Im} \; \widetilde g(\omega) =
\frac{\gamma \, \omega^3 \, \omega_0^2}{ 
\omega_0^4 \, \left( \omega^2 -  \omega_0^2 \right)^2 +
\gamma^2 \, \omega^6} \,.
\end{split}
\end{equation}
Inserting the expression 
$\Gamma_n(\omega) = \widetilde \Gamma_n(\omega)$ into Eq.~\eqref{20},
one obtains the length-gauge form for the imaginary part of the 
polarizability off resonance 
($\omega_0 \to \omega_{0n}$, $\gamma \to \Gamma_n$).

Quite surprisingly, the question of whether one should use
the length or velocity forms for the decay width off resonance,
i.e., in the interval $0 < \omega < \omega_{0n}$,
has not been answered conclusively in the literature.
It has often been stressed (e.g., in Ref.~\cite{Ko1978prl})
that the electric field strength $\vec E$ is a physical observable and thus
gauge invariant while the gauge-dependent vector potential 
$\vec A$ is not.
Lamb noted in footnote~88 on p.~268 of Ref.~\cite{La1952}
that the interpretation of the 
quantum mechanical wave function is only preserved in 
the length gauge with the dipole $\vec r \cdot \vec E$ interaction, 
because the kinetic momentum 
changes from $\vec p \to \vec p - e\, \vec A$ in the 
presence of a vector potential, and therefore the 
$\vec p$ operator in the quantum mechanical Hamiltonian 
of the atom cannot be interpreted any more as a
kinetic momentum if the vector potential is nonvanishing
(see also Chap.~XXI of Ref.~\cite{Me1962vol2} 
and Refs.~\cite{JeEvHaKe2003,EvJeKe2004}).
On the other hand, 
in Ref.~\cite{LaSoPlSo2002}, the authors explain in the
text after Eq.~(2) that ``the velocity [gauge]
form follows originally from the [fully relativistic] QED description''
and should therefore be used off resonance, for the obvious reason that
the nonrelativistic limit of the Dirac matrix vector $\vec\alpha$,
which enters the relativistic expression for the self-energy~\cite{Mo1974a}
is the momentum operator $\vec p/(m c)$.
While the length-gauge results seem to be 
generally preferred in the literature,
the use of the length versus velocity forms remains
controversial, and all numerical results below are therefore 
indicated for both velocity and length gauge;
further considerations on the choice of the gauge are beyond the scope
of the current article.

\begin{figure}[t!]
\includegraphics[width=0.8\linewidth]{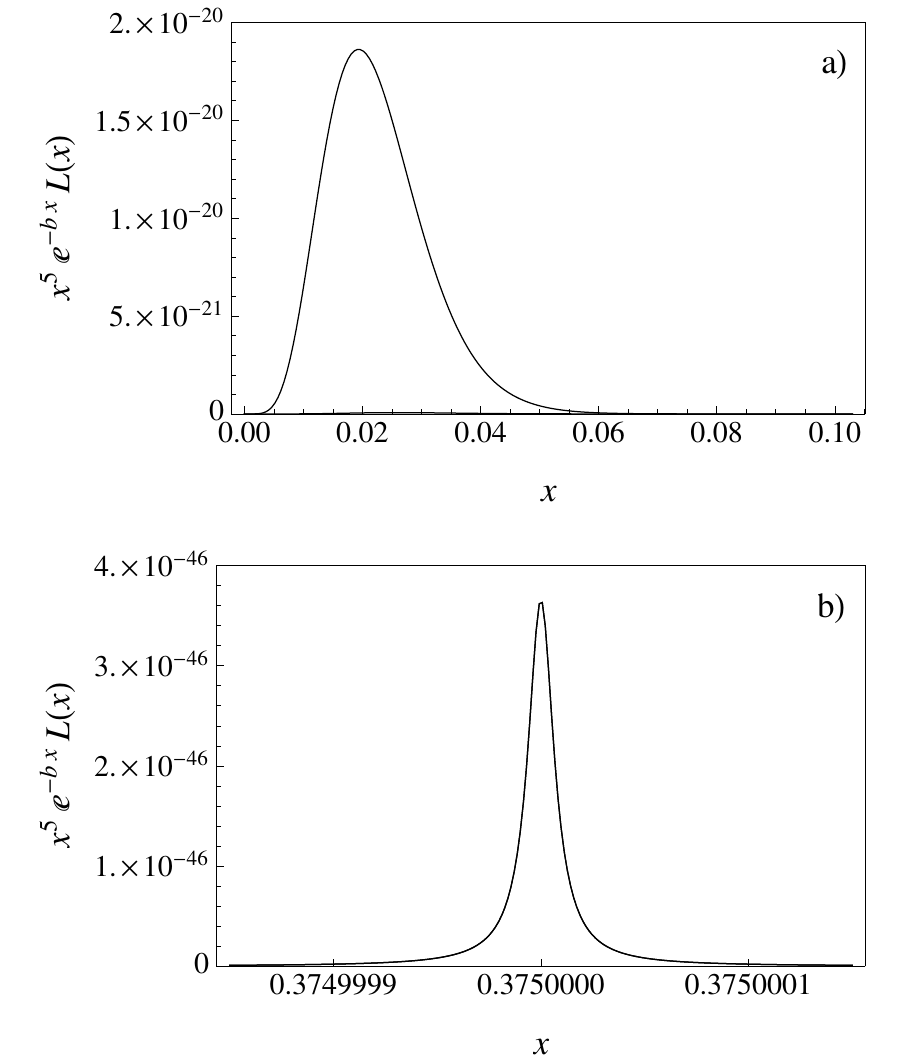}\\
\caption{\label{plots} 
Plot of the integrand $x^5 \, \ee^{-b\,x} \, L_s(x)$ 
defined in Eq.~\eqref{eq2} in the 
interval characteristic of the thermal peak 
$x \approx 5/b = 0.01583$ [Fig.~(a), parameter $s=1$] and near the resonant peak
$x \approx 0.375$ [Fig.~(b), indistinguishable curves for 
$s=1$ and $s=3$]. The thermal peak [Fig.~(a)]
yields the dominant 
contribution to the model integral $J$ defined in Eq.~\eqref{eq1}.}
\end{figure}

{\em Model example.---}In order to illustrate the 
numerical evaluation of Eq.~\eqref{force1}, we consider
a dimensionless model integral which is obtained 
by replacing frequency, transition width and $\beta$ with their
dimensionless equivalents:
\begin{align*}
 \omega &\to x=\hbar\,\omega/E_h,\hphantom{{}_0{}_0}\quad
 \Gamma \to \gamma=\hbar\,\Gamma/E_h,\quad\\
 \omega_0 &\to x_0=\hbar\,\omega_0/E_h,\quad
 \beta \to b=\beta\,E_h=E_h/(k_B T),
\end{align*}
where $E_h$ is the Hartree energy. The resulting integral 
\begin{subequations}
\begin{align}
\label{eq1}
J =& \; \int_0^\infty \dd x \, x^5 \, \ee^{-b\,x} \, L_s(x) \,,
\\[0.23ex]
\label{eq2}
L_s(x) =& \; 
{\rm Im}\left[ 1/\left(x_0 - x - \tfrac{\ii \gamma}{2} 
\big(\tfrac{x}{x_0}\big)^s\right) \right] \,,
\end{align}
\end{subequations}
contains the ``Boltzmann factor'' $\ee^{ -b\,x}$ which models the hyperbolic
sine in the denominator of the integrand of Eq.~\eqref{force1}, and the
imaginary part is taken for a single resonance  function that models the
Lorentzian line profile [with $s=1$ and $s=3$ for the analogues of
Eq.~\eqref{velgauge2} and Eq.~\eqref{lengauge2} respectively].  We choose the
temperature as $T = 1000\,K$, corresponding to $b=315.775$, and 
resonance parameters for the lowest (1S--2P) transition in the hydrogen
atom: $x_0 = \tfrac38$, and $\gamma =  1.5162\times 10^{-8}$.
The imaginary part of the Lorentzian ``polarizability term''
$L_s(x)$ is highly peaked near the 
resonance energy $x = x_0$ and for $s=0$, and the full width at half
maximum (FHWM) is  equal to $\gamma$.
The relative change of the prefactor $x^5 \, \exp(-b \, x)$
over the interval
$x \in (x_0 - \tfrac12 \, \gamma, x_0 + \tfrac12 \, \gamma)$
is smaller than $10^{-5}$.
One thus conjectures that the replacement
$L_s(x) \to \pi \, \delta(x - x_0) $
should provide for an excellent numerical approximation 
to the contribution of the Lorentzian peak to the 
integral $J$. Indeed,
\begin{align}
\label{peakterm}
P_1 =& \; \int\limits_{0.374999}^{0.375001}
\frac{\dd x\, x^5}{\ee^{b\,x}} \, L_s(x) =
\left\{
\begin{array}{l l}
8.669\times 10^{-54} & \;\textrm{($s=1$)}\\
8.669\times 10^{-54} & \;\textrm{($s=3$)}\\
\end{array}\right. \,,
\nonumber\\[0.23ex]
P_2 =& \; \int\limits_{0.374999}^{0.375001} 
\frac{\dd x\, x^5}{\ee^{b\,x}} \, \pi \, \delta(x - x_0) 
= 8.711 \times 10^{-54} \,,
\end{align}
confirming that $P_1 \approx P_2$ for the peak term.
However, this treatment ignores the possibility of 
far off-resonant driving of the transition for 
$x \ll x_0$. A numerical evaluation 
of the infrared thermal spectrum leads
to the result 
\begin{equation}
Q_s = \int\limits_{0}^{0.374999} 
\frac{\dd x\, x^5}{\ee^{b\,x}} \, L_s(x) =
\left\{
\begin{array}{l l}
3.741\times 10^{-22} & \quad \textrm{($s=1$)}\\
1.550\times 10^{-24} & \quad \textrm{($s=3$)}\\
\end{array}\right. ,
\end{equation}
which is larger than~\eqref{peakterm} by roughly 30~orders of magnitude.
The exact numerical result for $J$ fulfills $J_s \approx Q_s$ to six decimals
for $s=1$ and $s=3$; the effect is dominated by off-resonance absorption
(see also Fig.~\ref{plots}).

\begin{figure}[th!]
\begin{center}
\includegraphics[width=0.8\linewidth]{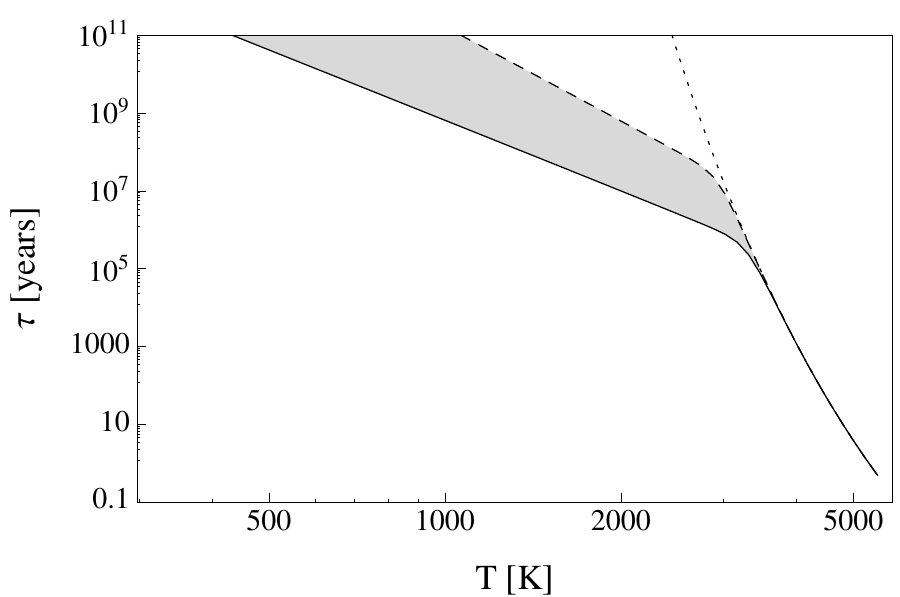}
\caption{\label{fig2} The characteristic slowdown time due to friction for
ground state atomic hydrogen as a function of the blackbody radiation
temperature.  The solid line shows the results using the dynamic
polarizability of Eq.~\eqref{20} in length-gauge form [Eq.~\eqref{lengauge2}],
the long-dashed line is the velocity-gauge form~\eqref{velgauge2},
the shaded area is in between,
and the short-dashed line results from Dirac $\delta$ peaks given in Eq.~\eqref{delta}.}
\end{center}
\end{figure}

\begin{figure}[th!]
\begin{center}
\includegraphics[width=0.8\linewidth]{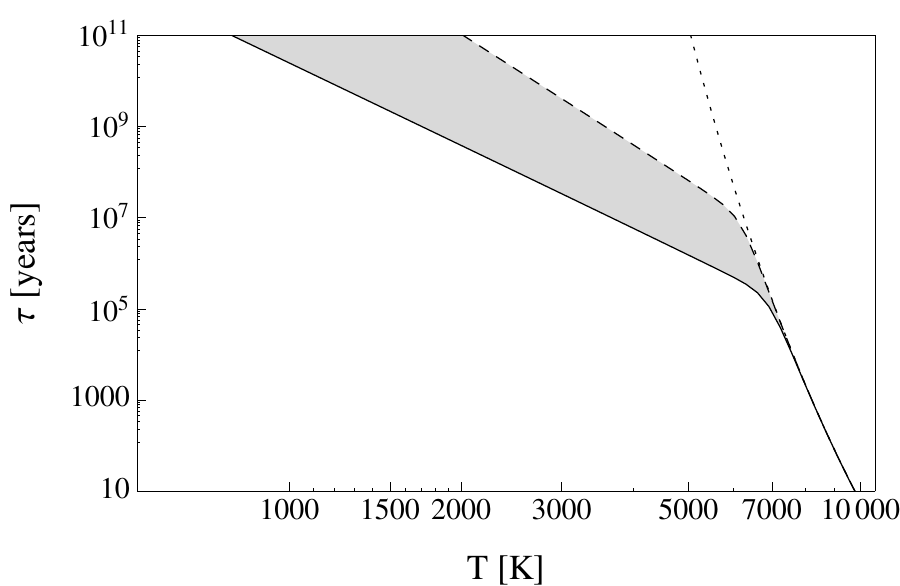}
\caption{\label{fig3} Same as Fig.~\ref{fig2}
for a ground state helium atom.}
\end{center}
\end{figure}

\begin{figure}[th!]
\begin{center}
\includegraphics[width=0.8\linewidth]{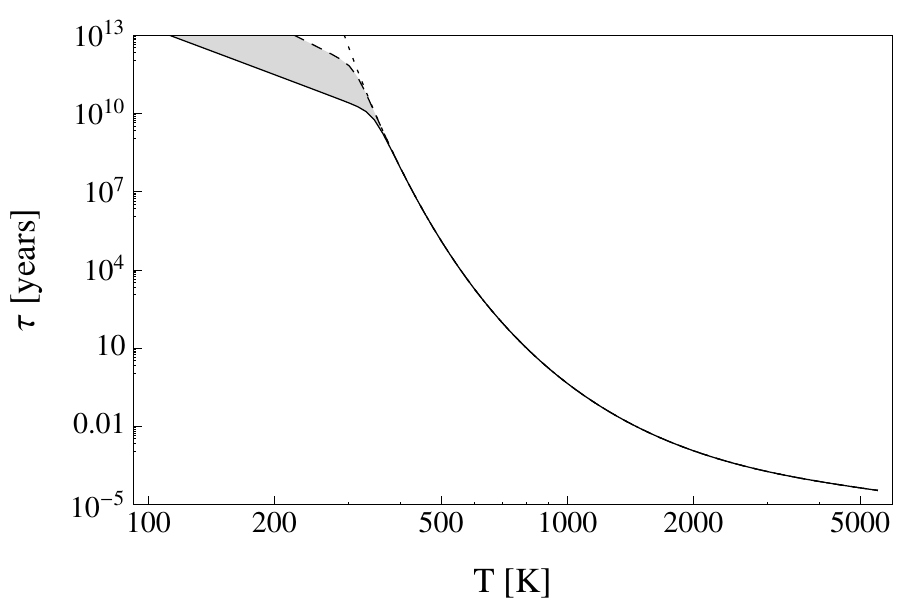}
\caption{\label{fig4} Same as
Figs.~\ref{fig2} and~\ref{fig3} 
for metastable helium ($2{^3}S_1$).}
\end{center}
\end{figure}

\begin{table}[th!]
\caption{Frequencies, oscillator strengths, lifetimes, and level widths for
transitions between the ground $1$S state and $n^{1}$P levels within the
hydrogen atom.}
\label{tableh}
\begin{center}
\begin{tabular}{cD{.}{.}{11}D{.}{.}{11}D{.}{.}{5}l}
\hline
\hline
\multicolumn{1}{c}{$n$} &
\multicolumn{1}{c}{$\omega_{0n}$ [a.u.]} &
\multicolumn{1}{c}{$f_{0n}$ [a.u.]} &
\multicolumn{1}{c}{$1/\Gamma_n$ [ns]} &
\multicolumn{1}{c}{$\Gamma_n$ [a.u.]}\\
\hline
2 & 0.375\,000\,000 & 0.416\,196\,717 &  1.595  & 1.5162\,10$^{-8}$ \\
3 & 0.444\,444\,444 & 0.079\,101\,562 &  5.268  & 4.5911\,10$^{-9}$ \\
4 & 0.468\,750\,000 & 0.028\,991\,029 & 12.346  & 1.9507\,10$^{-9}$ \\
5 & 0.480\,000\,000 & 0.013\,938\,344 & 23.949  & 1.0078\,10$^{-9}$ \\
\hline
\hline
\end{tabular}
\end{center}
\end{table}

\begin{table}[th!]
\caption{Same as Table~\ref{tableh}
for transitions between the ground $1^{1}$S state and the singlet $n^{1}$P levels
within the helium atom.}
\label{tablehesinglet}
\begin{center}
\begin{tabular}{cD{.}{.}{11}D{.}{.}{11}D{.}{.}{5}l}
\hline
\hline
\multicolumn{1}{c}{$n$} &
\multicolumn{1}{c}{$\omega_{0n}$ [a.u.]} &
\multicolumn{1}{c}{$f_{0n}$ [a.u.]} &
\multicolumn{1}{c}{$1/\Gamma_n$ [ns]} &
\multicolumn{1}{c}{$\Gamma_n$ [a.u.]}\\
\hline
2 & 0.779\,881\,291 & 0.276\,1647 & 0.555     & 4.3514\,10$^{-8}$ \\
3 & 0.848\,578\,015 & 0.073\,4349 & 1.728     & 1.3998\,10$^{-8}$ \\
4 & 0.872\,654\,727 & 0.029\,8629 & 3.975     & 6.0852\,10$^{-9}$ \\
5 & 0.883\,818\,387 & 0.015\,0393 & 7.653     & 3.1607\,10$^{-9}$ \\
\hline
\hline
\end{tabular}
\end{center}
\end{table}

\begin{table}[th!]
\caption{Same as Table~\ref{tablehesinglet}
for transitions from the triplet metastable $2^{3}$S state to triplet $n^{3}$P
levels (helium atom).}
\label{tablehetriplet}
\begin{center}
\begin{tabular}{cD{.}{.}{11}D{.}{.}{11}D{.}{.}{5}l}
\hline
\hline
\multicolumn{1}{c}{$n$} &
\multicolumn{1}{c}{$\omega_{0n}$ [a.u.]} &
\multicolumn{1}{c}{$f_{0n}$ [a.u.]} &
\multicolumn{1}{c}{$1/\Gamma_n$ [ns]} &
\multicolumn{1}{c}{$\Gamma_n$ [a.u.]}\\
\hline
2 & 0.042\,065\,187 & 0.539\,0861 & 98.202    & 2.4623\,10$^{-10}$ \\
3 & 0.117\,148\,294 & 0.064\,4612 & 98.154    & 2.4635\,10$^{-10}$ \\
4 & 0.142\,905\,024 & 0.025\,7689 & 142.886   & 1.6923\,10$^{-10}$ \\
5 & 0.154\,678\,191 & 0.012\,4906 & 225.643   & 1.0716\,10$^{-10}$ \\
\hline
\hline
\end{tabular}
\end{center}
\end{table}

{\em Simple Atoms.---}Returning from our model example~\eqref{eq1} to realistic
simple atoms, we provide results for the numerical integration of
Eq.~\eqref{force1} for hydrogen atoms in Fig.~\ref{fig2} and for helium atoms
in their ground and metastable triplet states in Figs.~\ref{fig3}
and~\ref{fig4}, respectively.  The friction force is expressed in terms of its
corresponding characteristic slowdown time $\tau=m\,v/F$.  For atomic hydrogen
and helium, the dynamic polarizability~\eqref{20} has been used with the
parameters listed in Tables~\ref{tableh}---\ref{tablehetriplet}. The input data
have been partially calculated by us, and the transition frequencies and
oscillator strengths have been verified against those given in
Refs.~\cite{Th1987,Dr2005}. The total decay rates
used in the calculation include the decays to both $^1$S and $^1$D states.  The
temperature at which the full Lorentz profile results start to deviate from the
Dirac-$\delta$ peaks is given by $T^* = \frac{\hbar \omega_{02}}{k_B x}$
where $x$ is the greater of the two real and positive (rather than complex)
solutions of the equation $x^7 \ee^{-x} = \frac{32}{21} \pi^5
\frac{\Gamma_2}{\omega_{02}}$ (velocity gauge) and $x^9 \ee^{-x} =
\frac{128}{15} \pi^7 \frac{\Gamma_2}{\omega_{02}}$ (length gauge, $n=2$ is the
principal quantum number of the lowest excited state). For an equation of the
form $x^n \, \ee^{-x} = A$, this particular solution can be expressed as $x =
-n \, W_{-1}(-A^{1/n}/n)$, where $W$ is the generalized Lambert $W$
function~\cite{CoGoHaJeKn1996}.  In velocity gauge, $T^*$ evaluates to
$3293 \, \K$ for hydrogen, 
$6927 \, \K$ for singlet and $346 \K$ for triplet helium. 
In length gauge we have 
$T^* = 2954\, \K$ for hydrogen, 
$6208 \, \K$ for singlet and $312 \, \K$ for triplet helium 
(confirmed in Figs.~\ref{fig3} and~\ref{fig4}).

{\em Conclusions.---}In this Letter, we show that far off-resonant driving of
atomic transitions yields the dominant contribution to the blackbody friction
force on moving atoms, due to the overlap of the infrared tail of the Lorentzian
profile with the infrared thermal peak of the blackbody radiation.
It is thus imperative to take the finite lifetime of the atomic
resonances and their corresponding width into account. Numerical results for
simple atoms are provided in Figs.~\ref{fig2}---\ref{fig4}.  The feasibility
of an experimental verification of the predictions of this Letter remains to be
studied.  Of the atomic systems considered here, the largest effect is expected
for the metastable $^3$S$_1$ state in helium.  In this case and for a
temperature equal to the melting point of tungsten (3695K) the characteristic
slowdown time is computed to be 3016s ($\approx50$ minutes), which makes the
friction effect difficult to observe in laboratory experiments, but perhaps not
impossible. The general importance of an accurate understanding of blackbody friction 
for astrophysical processes has already been stressed in Ref.~\cite{MkPaPoSa2003}.
Further remarks on conceivable astrophysical consequences of the calculations
reported here are beyond the scope of this Letter.

{\em Acknowledgments.---}This project was supported by NSF, DFG
and a precision measurement grant (NIST).

\end{document}